\documentclass{article}

\PassOptionsToPackage{numbers, compress}{natbib}

\usepackage[main, final]{neurips_2026}

\makeatletter
\renewcommand{\@toptitlebar}{\vskip 0in}
\renewcommand{\@bottomtitlebar}{\vskip 0in}
\renewcommand{\@notice}{}
\makeatother

\usepackage[utf8]{inputenc} 
\usepackage[T1]{fontenc}    
\usepackage{amsmath}        
\usepackage{amssymb}        
\usepackage{hyperref}       
\usepackage{url}            
\usepackage{booktabs}       
\usepackage{amsfonts}       
\usepackage{nicefrac}       
\usepackage{microtype}      
\usepackage{xcolor}         
\usepackage{graphicx} 
\usepackage{multirow}

\title{Reinforcement learning to improve large language model-based automated code compliance systems}

\author{%
  \textbf{Jack Wei Lun Shi}\textsuperscript{\textmd{1}} \quad
  \textbf{Minghao Dang}\textsuperscript{\textmd{1,2}} \quad
  \textbf{Wawan Solihin}\textsuperscript{\textmd{1,3}}
  \quad
  \textbf{Justin K.W. Yeoh}\textsuperscript{\textmd{1}} \\[2ex]
  \normalfont\normalsize
  \textsuperscript{1}National University of Singapore \quad
  \textsuperscript{2}Harbin Institute of Technology \\[1ex]
  \normalfont\normalsize
  \textsuperscript{3}NovaCITYNETS
}

\begin{document}

\maketitle

\begin{abstract}
    Large language model (LLM)-based approaches for automated code compliance
    (ACC) of building regulations are prone to generating incorrect and
    hallucinated computer-processable rules. This paper introduces P4IR, a
    two-stage framework that uses supervised fine-tuning (SFT) to instill
    domain knowledge in an LLM, followed by Group Relative Policy Optimization (GRPO) to improve the accuracy of the generated intermediate
    representations in the form of high-level code skeletons. The framework achieved final reductions of up to 21.3\% and 41.0\% in tree edit distance and token-level Levenshtein distance respectively, relative to the SFT baselines. Comparative analysis demonstrates that this approach in a zero-shot setting outperforms leading LLMs in both code structure and semantics, evaluated via few-shot prompting. Additionally, the GRPO stage produced a statistically significant reduction in false positives. By combining SFT with GRPO to optimize directly for domain-specific objectives, this approach offers a path toward more accurate and reliable LLM-based ACC systems.
\end{abstract}

\begin{quote}
\end{quote}

\section{Introduction}

Compliance with building regulations is crucial for ensuring the safety,
functionality, and longevity of structures, while directly influencing
project timelines and costs. Singapore's Building and Construction
Authority forecasts construction demand to reach S\$53 billion in 2025,
representing an 11.7\% increase over pre-pandemic levels, and to remain
robust through 2029, averaging between S\$39 billion and S\$46 billion
annually \cite{ref1}. In this dynamic environment, automated code compliance
(ACC) systems can drastically reduce design review times, improve
accuracy, and enhance overall project efficiency within the
Architecture, Engineering, Construction, and Operations (AECO) industry.
Despite advancements in ACC systems, the complexity of translating
building regulatory clauses into computer-processable formats remains a
significant challenge \cite{ref2}. Conventional approaches often rely on
time-consuming and hard-coded rule-based methods, highlighting the need
for more flexible and efficient solutions. To manage this complexity,
building regulations are often first translated into Intermediate
Representations (IRs), which are structured representations that can
better support the generation of computer-processable formats. In the
ACC domain, such representations have been used to capture hierarchical
information and semantic rules \cite{ref3,ref4}.

With the recent advances in artificial intelligence (AI), large language
models (LLMs) have emerged as a promising approach to generate
structured and coherent representations from unstructured regulatory texts.
Early work by Zhang \cite{ref5} explored ChatGPT for converting
natural-language building code requirements into executable Python
logic, demonstrating that the model generates regulatory-compliant code
with high efficiency and showcasing its potential to accelerate ACC
implementation. Fuchs et al. \cite{ref6} showed that GPT-3.5, in a few-shot
setting with as few as five exemplars and carefully crafted system
prompts, can translate building regulations into compact IRs using
LegalRuleML, achieving F1-scores of approximately 70\%. These results
were further improved by representative exemplar sampling and
self-consistency strategies.

Despite these promising results, LLMs can be prone to hallucination
\cite{ref7} (i.e., generating outputs that are statistically plausible but
flawed). Supervised fine-tuning (SFT) can inadvertently amplify this
phenomenon by forcing the model to fabricate responses when instructions
exceed its knowledge boundaries, as it trains the model to always
provide an answer rather than express uncertainty \cite{ref8}. Consequently,
the LLM outputs, such as the IRs, are not always reliable for rule
interpretation \cite{ref9}. Even with few-shot prompting, recent
state-of-the-art (SOTA) models struggle to generate IRs that match the
ground truth in both structure and semantics.

Reinforcement learning (RL) offers a promising approach, as it optimizes
explicit reward functions and can therefore be aligned with
domain-specific objectives, which has been shown to reduce
hallucinations in practice \cite{ref10,ref11}. Among various RL-based
algorithms, Group Relative Policy Optimization (GRPO) reduces
computational and memory requirements by eliminating the critic model
while also stabilizing training \cite{ref12}.

This paper proposes P4IR (\textbf{P}olicy Optimization \textbf{for}
\textbf{I}ntermediate \textbf{R}epresentations), a two-stage GRPO-based
ACC framework. The first stage employs SFT to equip a pre-trained LLM
with domain knowledge to establish semantic accuracy \cite{ref13}. The
second stage then applies GRPO to improve the structure of generated IRs
(hereafter ``code skeletons'') and reduce hallucinations. This
reinforcement stage is guided by an accuracy reward function based on
Jaccard similarity \cite{ref14}, a metric measuring the alignment between
the generated and reference code skeletons by comparing their classes,
functions, and parameters. Thereafter, a systematic evaluation
investigates the influence of initial SFT maturity on subsequent GRPO
performance. The paper also benchmarks P4IR against other LLMs,
providing empirical evidence that the two-stage framework yields better
results than larger general-purpose models evaluated with few-shot
prompting.

\section{Literature review}
ACC refers to the evaluation of a Building Information Model (BIM)
against a computable representation of building regulations. The overall
structure of an ACC system generally consists of four stages: rule
interpretation, building model preparation, rule execution stage, and
reporting of checking results. Within this ACC system, the primary
challenge lies in the initial rule interpretation.

Over the past decades, significant efforts have been dedicated to
improving ACC systems. These include using semantic decomposition
methods to convert textual regulations into rule execution languages
without requiring programming expertise \cite{ref16}, establishing
ontology-based methods \cite{ref17} and domain-specific languages \cite{ref18}
to formalize regulatory knowledge within BIM workflows, and adopting
visual programming languages to generate transparent process diagrams
that allow rule experts to build and interpret compliance logic visually
\cite{ref19}. However, these conventional methods share limitations. They
struggle with the inherent ambiguity and implicit knowledge within
regulatory texts, demanding manual effort to interpret and translate
rules into computer-processable formats. Furthermore, many of these
systems are characterized by hard-coded implementations that lack the
flexibility to adapt to evolving regulations or to generalize
effectively across diverse and complex rule sets.

\subsection{Natural language processing-based approaches in ACC}

To overcome these challenges, research has moved toward harnessing
natural language processing (NLP) methods to support various ACC
processes. These NLP applications can be broadly classified into five
interconnected areas: (1) information extraction of structured
requirements from textual regulations, (2) formalizing these
requirements into computer-processable rules, (3) constructing knowledge
representations of building regulations, (4) executing compliance checks
against BIM, and (5) creating question-answering (QA) support systems.
These classifications are not mutually exclusive, and many research
initiatives adopt a hybrid approach, integrating techniques from several
of these categories to create more comprehensive ACC solutions.

In the area of information extraction, Zhang and El-Gohary \cite{ref20}
presented a semantic rule-based NLP method that automatically extracts
regulatory requirements from construction codes using pattern matching
and ontological semantics. Their approach achieved a precision and
recall of 0.969 and 0.944 respectively when tested in extracting of
quantitative requirements from the 2009 International Building Code
(IBC). Song et al. \cite{ref21} introduced a deep learning-based NLP
framework for extracting predicate argument structures from building
design rule sentences and converting them into machine-readable
representations for BIM-enabled compliance verification. Once this
information is extracted, the next step is to convert the fragments into
programmatic rules, a task known as semantic parsing. For example, Guo
et al. \cite{ref22} transformed the extracted terms and relations into
SPARQL using term matching and semantic similarity analysis.

Parallel to these efforts in parsing, other work has focused on
constructing comprehensive knowledge representations. For example, Zheng
et al. \cite{ref23} proposed a knowledge-informed framework that uses NLP to
construct a domain ontology of building concepts, enrich BIM with
semantic metadata, and align regulatory text to ontology entities. Their
approach enabled the automatic generation of SPARQL queries for rule
interpretation and compliance checking, achieving 90.1\% accuracy in
semantic alignment and a fivefold increase in interpretation speed.
Similarly, Peng and Liu \cite{ref24} presented an NLP-driven approach that
preprocesses building regulations using the NLPIR Chinese word
separation system and an attention-based conditional random fields (CRF)
model to extract and classify domain entities, attributes, and their
relationships. They constructed a semantic knowledge graph integrated
with BIM elements to support downstream automated compliance checking,
and validated the system on four engineering projects with a review
accuracy exceeding 96\%. Integrating these methods, ontology-based
approaches have also been used for ACC by formalizing structural fire
safety requirements into an ontology that integrates diverse regulatory
texts and expert insights to enable automatic inference of compliance
checks \cite{ref17} and by deploying an ontology-informed information
extraction framework to convert building energy conservation codes into
machine-readable compliance rules \cite{ref25}.

For the purposes of executing compliance checks, Zhang and El-Gohary
\cite{ref26} developed SNACC, a unified ACC system that leverages semantic
NLP to automatically extract and transform regulatory requirements into
logic rules, aligns these with BIM-derived logic facts, and applies
logic reasoning to perform fully automated compliance verification.
Their system achieved 98.7\% recall and 87.6\% precision in
non-compliance detection regarding Chapter 19 of the 2009 IBC. Wu et al.
\cite{ref27} extended SNACC by introducing the I-SNACC framework, which
combines invariant signature-based fact extraction with logic reasoning,
and adds modules for semi-automated rule expansion, activation condition
generation, and interactive model validation to further automate
building code compliance checks with minimal user input. Lastly, NLP has
also been used to create QA systems for efficient information retrieval
of building regulatory information and from BIM. For example, Wang et
al. \cite{ref28} presented a QA system that achieved an accuracy score of
81.9\% using NLP for information extraction directly from BIM and IFC
models, allowing users to ask questions and receive answers in natural
language.

\subsection{Transformer- and LLM-based approaches in ACC}

With the advent of transformer-based LLMs, ACC research has begun to
look beyond traditional NLP approaches for state-of-the-art methods.
Transformer-based approaches are also implemented across the previously
mentioned five areas. Respectively, Okonkwo \cite{ref29} evaluated
the performance of word embeddings and transformer-based models (e.g.,
Bidirectional Encoder Representations from Transformers (BERT) and
Sentence BERT) for automated information extraction of semantic
regularities from unstructured British building regulations, achieving
up to 80\% word-level and perfect sentence-level accuracy. Additionally,
Zheng et al. \cite{ref30} developed a domain-specific BERT model to classify
the machine interpretability of building regulatory clauses, filtering
out non-computable text to enhance rule interpretation. Fuchs et al.
\cite{ref4} leveraged transformer-based semantic parsers augmented with
structured intermediate formats, showing that introducing reversible
representations and a hierarchical parsing step can cut training time by
nearly 75\% and boost F1-scores by up to 6.6\%, while also offering an
interpretable, human-in-the-loop translation process.

For rule interpretation, Zheng et al. \cite{ref2} presented LLM-FuncMapper,
which builds a library of 66 atomic functions for low and high order
logic in building codes and uses an LLM to identify predefined
functions. When employed with few-shot methods, LLM-FuncMapper retrieved
81.55\% of the relevant atomic functions within its top five
recommendations, surpassing that of a fine-tuned BERT model. To
construct knowledge representations, Iranmanesh et al. \cite{ref31}
transformed IFC building data into graph-based formats by retrieving
building object properties and their relations using an LLM-based graph
retrieval-augmented generation technique. For LLM-based QA systems on
building regulations, Xue et al. \cite{ref32} paired a BM25-based sparse
retriever with a reader built on fine-tuned and distilled LLMs, while
Zhong et al. \cite{ref33} used a fine-tuned BERT model to analyze retrieved
regulatory passages.

The outputs of the preceding studies, whether formal ontologies,
queries, or programmatic functions, implicitly serve as IRs that bridge
complex building regulations and the final, computer-processable rules.
The quality of these IRs is therefore pivotal to the success of ACC
systems. To automate their generation, recent studies have adopted
advanced approaches such as SFT and LLM agents. For example, Fuchs et
al. \cite{ref9} employed SFT on T5 and BART models to translate regulations
directly into a formal IR in LegalRuleML, while Ying and Sacks \cite{ref34}
leveraged an agentic workflow to dynamically create intermediate outputs
(i.e., explicit reasoning traces and intermediate JSON files) for
orchestrating the compliance check. However, Fuchs et al. \cite{ref9} found
that such outputs can be flawed, with underspecified logic statements and
hallucinated object references. Similarly, Wu et al. \cite{wu2027structured}
identified a mismatch between standard cross-entropy training and the
structural consistency and logical coherence required for regulatory IRs,
highlighting the need for task-specific optimization objectives.
Consequently, while SFT can capture
latent concepts in building regulations and instill domain-specific
knowledge in the LLM \cite{ref13}, it does not by itself ensure that the
generated IRs are well-structured. This motivates a post-SFT
optimization stage that optimizes directly for output structure.

\subsection{Reinforcement fine-tuning}

Reinforcement fine-tuning (RFT) has become a popular strategy for
aligning LLMs with desired behaviors. Recently, Shao et al. \cite{ref12}
proposed GRPO as an efficient RFT algorithm, a variant of PPO that
foregoes the separate critic, instead estimating the baseline (i.e., the
expected return) from the average reward of a group of outputs sampled
for the same prompt, thereby eliminating the need for expensive value
function approximation. These simplifications yield more efficient and
stable training without sacrificing performance. For example, applying
GRPO-based RFT to a math-focused LLM yielded gains on mathematical
reasoning tasks, which often require generating a coherent
chain-of-thought to solve a problem. This increased accuracy on the
GSM8K math benchmark from 82.9\% to 88.2\%, while reducing training
resource requirements by removing the critic model \cite{ref12}.

However, despite its success in STEM fields, the efficacy of GRPO for
the domain-specific challenge of refining code skeletons for ACC has not
yet been investigated. Additionally, when applied to the ACC domain, the
effectiveness of GRPO for refining code skeletons is contingent on the
knowledge of the pre-trained LLMs. Since general-purpose models are not
pre-trained on specialized ACC datasets, they lack the domain-specific
knowledge required for nuanced rule interpretation, limiting the
effectiveness of applying GRPO directly to this task. This motivates
using an initial SFT stage to establish the domain-specific knowledge
and naming conventions before applying GRPO. Yet,
the interaction between this SFT stage and the subsequent GRPO stage has
not been systematically examined in ACC.

\subsection{Research gaps and objectives}

General-purpose LLMs, despite rapid benchmark gains, may not necessarily
transfer to highly specialized domains. Compact and domain-optimized
models such as BioMistral \cite{ref38}, FinBERT \cite{ref39}, and DeepSeekMath
\cite{ref12} have rivaled or surpassed far larger general-purpose models in
medicine, finance, and mathematics. This gap is more acute for ACC,
whose rule-conversion datasets are largely proprietary and absent from
the public corpora those models draw on, so relying on off-the-shelf
SOTA models is insufficient and motivates a framework that explicitly
aligns the model with the domain.

Within this landscape, even after domain-specific fine-tuning,
LLM-generated IRs can still contain underspecified logic and
hallucinated outputs. Two research gaps remain in addressing this.
First, standard SFT optimizes token-level prediction without explicitly
rewarding the accuracy of the generated IR structure \cite{wu2027structured}.
While GRPO presents a promising solution for addressing this gap through
task-specific rewards, its effectiveness has not been systematically
investigated for refining IRs in the ACC domain. Second, since the
domain-specific nature of ACC requires GRPO to build upon an LLM
previously fine-tuned with specialized knowledge, the interplay between
initial knowledge instillation via SFT and subsequent GRPO-based
refinement remains unexplored.

To address these gaps, this paper proposes a two-stage framework for
improving the accuracy of generated IRs. In the first stage, a
pre-trained LLM is fine-tuned on a domain-specific dataset, instilling
the model with the foundational knowledge needed to generate an initial
IR. In the second stage, GRPO is applied using a primary accuracy reward
function, together with syntactic and length penalties that discourage
malformed or trivially short outputs. This paper addresses the following
research questions: (1) Can GRPO further improve the accuracy of
generated IRs beyond what SFT achieves alone? (2) How does the degree of
initial SFT maturity influence the effectiveness of the subsequent GRPO
refinement stage? (3) How does P4IR, in a zero-shot setting, compare to
existing SOTA LLMs evaluated via few-shot prompting? By answering these
questions, this paper seeks to establish guideline on integrating SFT
and GRPO to generate high-quality and structure-correct IRs for ACC.

\section{Methodology}

This paper introduces P4IR (\textbf{P}olicy Optimization \textbf{for}
\textbf{I}ntermediate \textbf{R}epresentations), a framework for
improving the accuracy of generated code skeletons for ACC. A code
skeleton is the IR used in this paper, a high-level set of class and
function definitions for converting building regulations into
computer-processable rules, with low-level implementation details
omitted, as shown in Fig.~\ref{fig:1}. While Shi et al.\ \cite{ref13}
focused on generating only the main \texttt{CheckRule} function, this
paper considers the code skeleton of the full script, including its
supporting classes and functions and their hierarchical relationships.
GRPO is then applied to improve the accuracy of this IR. The low-level
details, such as the
retrieval and referencing of BIM entities and identifiers (e.g., GUID),
are extensive and rule-specific, making them challenging for LLMs to
learn. Separating them from the structure lets rule experts and
programmers add the implementation afterward, so a code skeleton
generated by P4IR serves as a starting point that accelerates the
development of complete compliance-checking scripts.

\subsection{Code skeleton as an intermediate representation}
\label{sec:dataset}

\begin{figure}[ht]
\centering
\includegraphics[width=\linewidth]{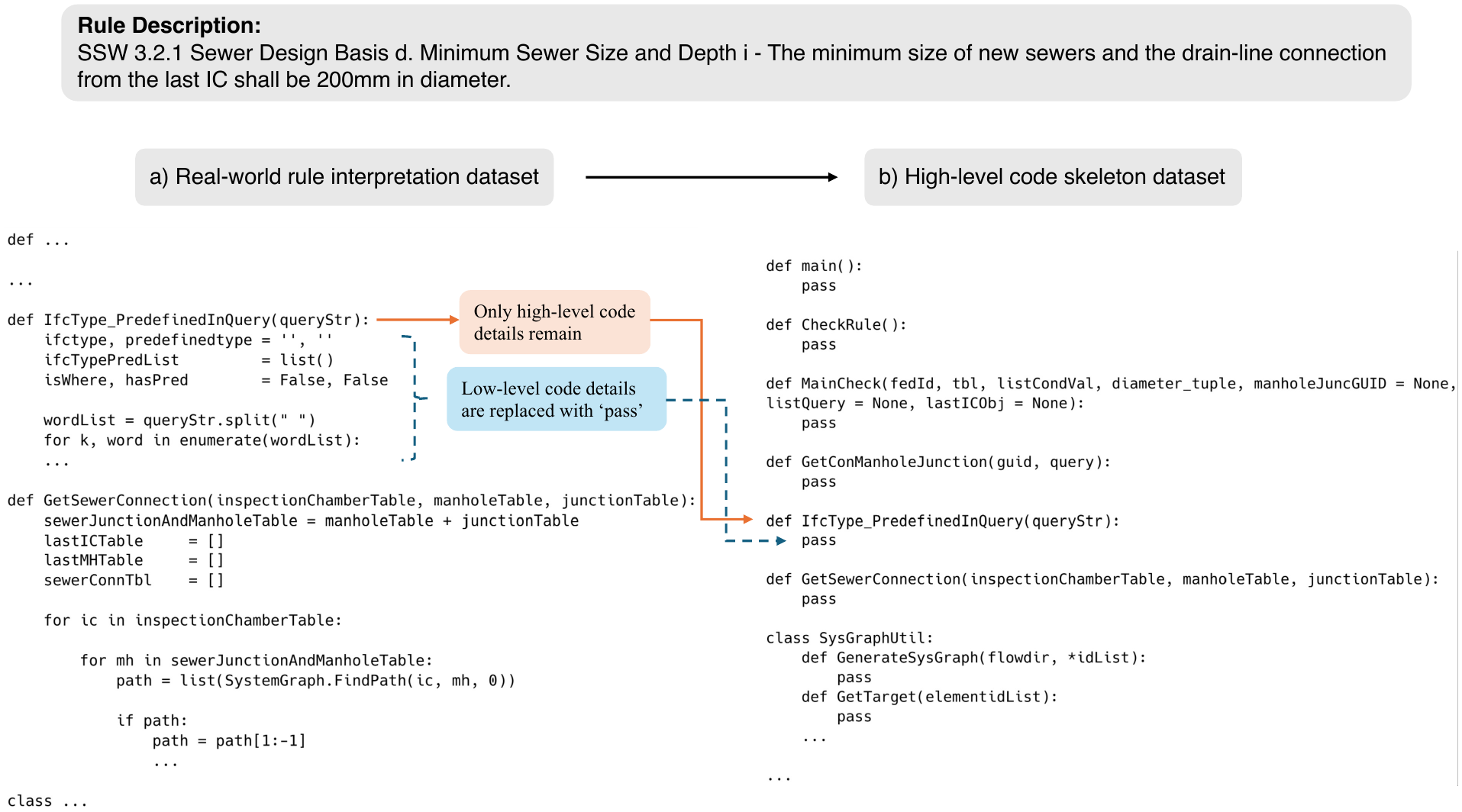}
\caption{Extracting the code skeleton from the rule interpretation dataset.}
\label{fig:1}
\end{figure}

The dataset used for training and evaluating P4IR is derived from a
real-world rule interpretation project described in Shi et al.\ \cite{ref13}.
To tailor it for the refinement
task in this paper, a preprocessing pipeline was applied. The dataset
contained pairs of regulatory text and their corresponding fully
implemented Python scripts, as shown in Fig.~\ref{fig:1}a. The regulatory texts
come from several Singapore codes of practice, such as Sewerage and
Sanitary Works (SSW) and Surface Water Drainage (SWD) code of practice
from Public Utility Board (PUB), and the Fire Code from Singapore Civil
Defence Force (SCDF).

Since this is a real-world project, the corresponding Python scripts are
tested and fully executable, and are therefore treated as ground truth.
From each script, only class and function definitions were extracted
(i.e., their names, parameters, and hierarchical nesting as shown in
Fig.~\ref{fig:1}). The low-level implementation details within each function body
were replaced with a stub (i.e., pass), as shown in Fig.~\ref{fig:1}b.

This process transformed each full script into a code skeleton, which
serves as the ground truth for the task. The abstraction focuses the
model on learning the structure of the code skeleton rather than the
details of its implementation, since generating fully implemented
scripts would carry a higher risk of hallucinating plausible but
incorrect low-level code. Following preprocessing for code-skeleton
generation, the resulting 732 regulation-skeleton pairs were partitioned
using a new random split into 664 training pairs and 68 test pairs.
Duplicate pairs were removed before splitting. Both sets contain
regulations from the same codes, and therefore some samples share
regulation texts or reference skeletons, reflecting common terminology
and implementation patterns in practice. The dataset spans a range of
complexity, with individual code skeletons containing anywhere between
0 and 28 classes and between 2 and 80 functions, including methods,
and nesting depths ranging from one to four levels. This allows the
model to be trained and evaluated on both simple and hierarchically
complex code skeletons.

\subsection{P4IR framework}

\begin{figure}[ht]
\centering
\includegraphics[width=\linewidth]{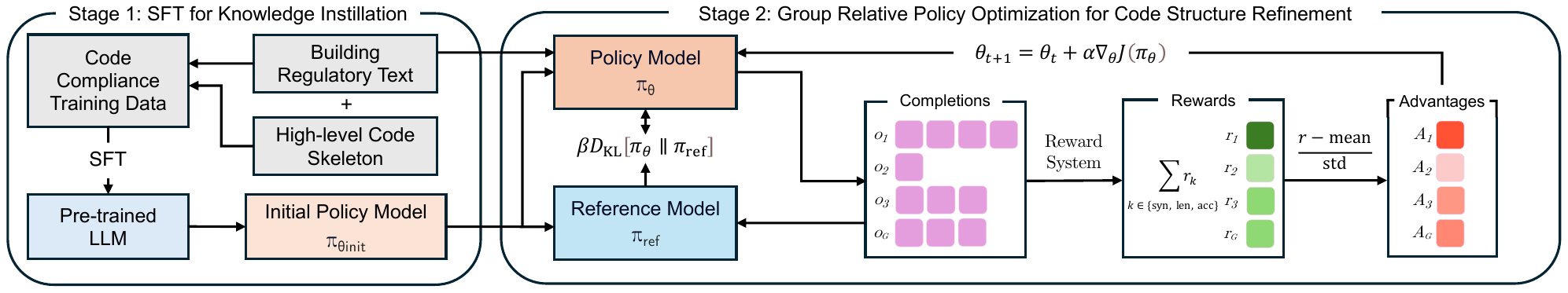}
\caption{Overall pipeline of the proposed framework.}
\label{fig:2}
\end{figure}

The Mistral 7B base model undergoes SFT on a real-world dataset of
Singapore building regulations and their corresponding code skeletons,
developed in collaboration with industry stakeholders. This fine-tuned
model is further optimized using GRPO, where accuracy and formatting
reward functions guide the model to reduce hallucinations and generate
accurate code skeletons. The overall pipeline of the framework is
depicted in Fig.~\ref{fig:2}.

The first stage of the framework, shown on the left portion of Fig.~\ref{fig:2},
uses SFT to equip the base LLM with knowledge of the ACC domain. The
objective of this stage is to train the model to map unstructured
regulatory text directly to its corresponding code skeleton. The
training dataset consists of pairs (\emph{q, y}), where \emph{q} is the
input regulatory text (i.e., query) and \emph{y} is the code skeleton.
This fine-tuning optimizes the model's parameters \(\theta\) by
maximizing the conditional likelihood of generating the target code
skeleton sequence \emph{y} given the input query \emph{q}, achieved by
minimizing the standard cross-entropy loss \(\mathcal{L}_{SFT}\) over
the training dataset \(\mathcal{C}\):

\begin{equation}
\mathcal{L}_{\mathrm{SFT}} = -\,\mathbb{E}_{(q,y)\sim \mathcal{C}} \sum_{t=1}^{|y|} \log \pi_{\theta}\!\left(y_t \mid q,\, y_{<t}\right)
\label{eq:sft}
\end{equation}

where \(\pi_{\theta}\) is the token distribution of the LLM and
\(y_{t}\) is the \emph{t}-th token of the target code skeleton. The
resulting model from this stage, denoted as \(\pi_{\theta init}\),
possesses the specialized knowledge of ACC domain conventions to
generate code skeletons and serves as the initial policy for the
subsequent GRPO refinement stage.

Next, the second stage of the framework, shown on the right portion of
Fig.~\ref{fig:2}, uses GRPO to refine the initial SFT policy \(\pi_{\theta init}\)
to improve the accuracy of its outputs. For a given regulatory text
input \emph{q}, the refinement process begins by sampling a group of
\emph{G} candidate code skeletons
\(\text{\{}o_{1},o_{2},o_{3},\ldots,o_{G}\text{\}}\) from the old policy
\(\pi_{\theta_{old}}\). Each output \(o_{i}\) is evaluated against a set
of rule-based reward functions (refer to Section~\ref{sec:rewards}) to obtain the
rewards \(\text{\{}r_{1},r_{2},r_{3},\ldots,r_{G}\text{\}}\). By
computing the mean and standard deviation of these rewards for
normalization, GRPO determines the relative advantages
\(\text{\{}A_{1},A_{2},A_{3},\ldots,A_{G}\text{\}}\) (i.e., the relative
quality of these outputs), defined as:

\begin{equation}
A_i = \frac{r_i - \mathrm{mean}\{r_1,\dots,r_G\}}{\mathrm{std}\{r_1,\dots,r_G\}}
\label{eq:advantage}
\end{equation}

where \(A_{i}\) represents the relative quality of the \emph{i}-th
output compared to others in the sampled group. The model is then
reinforced to favor outputs with a high advantage by updating the policy
parameters \(\theta\) through the optimization of the GRPO objective
function \cite{ref37}:

\begin{equation}
\begin{aligned}
\mathcal{J}_{\mathrm{GRPO}}(\theta) ={} & \mathbb{E}\Bigg[ \frac{1}{G}\sum_{i=1}^{G} \min\Bigg( \frac{\pi_{\theta}(o_i \mid q)}{\pi_{\theta_{\mathrm{old}}}(o_i \mid q)}\, A_i, \\
& \mathrm{clip}\!\left( \frac{\pi_{\theta}(o_i \mid q)}{\pi_{\theta_{\mathrm{old}}}(o_i \mid q)},\, 1-\epsilon,\, 1+\epsilon \right) A_i \Bigg) \\
& - \beta_{\mathrm{KL}}\, \mathcal{D}_{\mathrm{KL}}\!\left[ \pi_{\theta}(o_i \mid q) \,\big\|\, \pi_{\mathrm{ref}}(o_i \mid q) \right] \Bigg]
\end{aligned}
\label{eq:grpo}
\end{equation}

\(\mathbb{E}\) denotes the expectation over the sampled queries \emph{q}
and the corresponding group of generated outputs \({\{ o}_{i}\}\). The
policy ratio
\(\frac{\pi_{\theta}\left( o_{i} \middle| q \right)}{\pi_{\theta_{old}}\left( o_{i} \middle| q \right)}\)
compares the probability of generating an output \(o_{i}\) under the
current policy \(\pi_{\theta}\) versus the old policy
\(\pi_{\theta_{old}}\), while \(\epsilon\) and \(\beta_{KL}\) are
hyperparameters. The clipping function limits the policy ratio to a
small interval \(\lbrack 1 - \epsilon,\ 1 + \epsilon\rbrack\),
preventing excessively large policy updates and ensuring training
stability. The final term is a KL divergence penalty, scaled by
\(\beta_{KL}\), that regularizes the policy, ensuring it does not
deviate excessively from the knowledge of the reference SFT policy
\(\pi_{ref}\). By optimizing this objective, the framework uses its
reward functions to iteratively guide the policy toward generating more
accurate code skeletons.

\subsection{Reward design for refining code skeletons}
\label{sec:rewards}

\begin{figure}[ht]
\centering
\includegraphics[width=\linewidth]{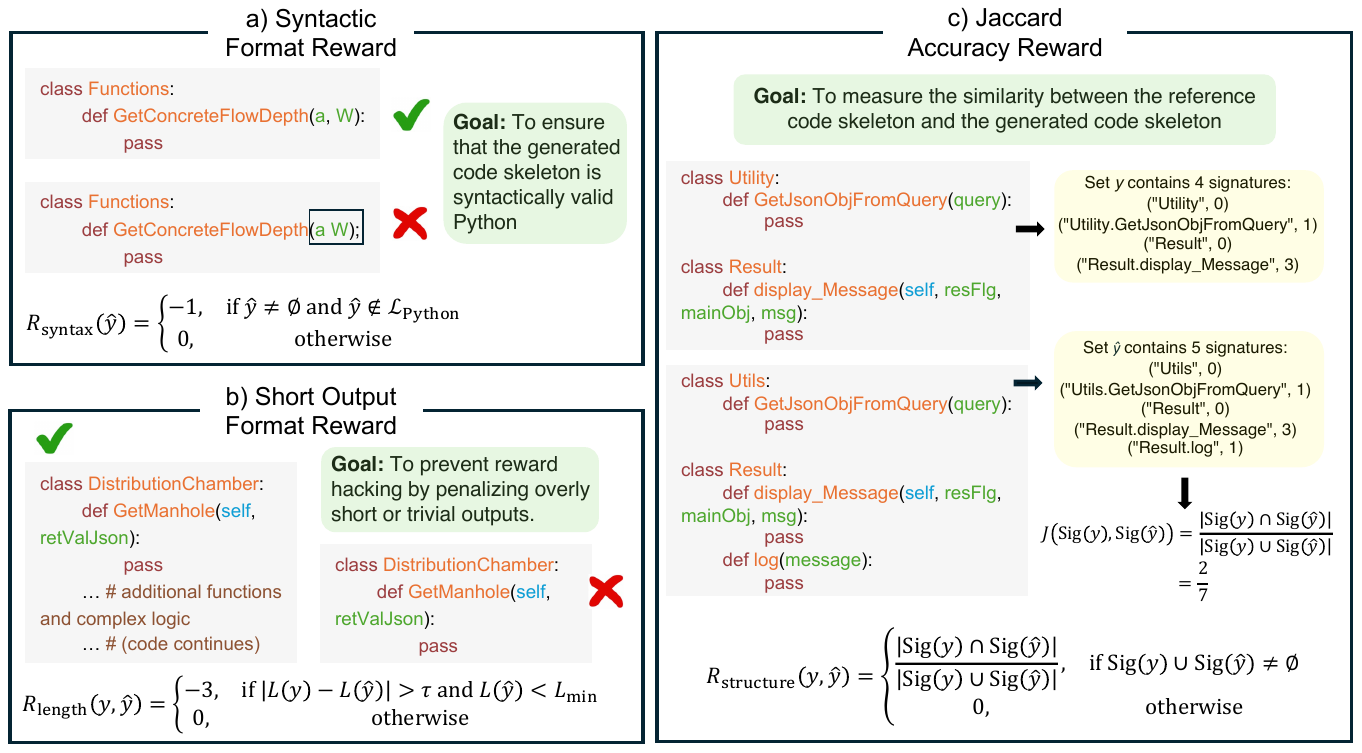}
\caption{Reward system for the proposed framework.}
\label{fig:3}
\end{figure}

The reward system provides the optimization signal that guides the LLM's
policy toward generating more accurate code skeletons. As shown in Fig.~\ref{fig:3}, it comprises multiple reward functions that evaluate each generated
skeleton, denoted \(\widehat{y}\), against its ground truth \(y\). It
combines a primary accuracy reward with two formatting penalties
(hereafter also referred to as ``rewards'') that together target both
the structure and the syntactic validity of the output.

The accuracy reward, denoted \(R_{\text{structure}}\) as shown in Fig.~\ref{fig:3}c, measures the alignment between the generated and reference code
skeletons. It is formulated as the Jaccard similarity between the sets
of hierarchical signatures extracted from the two code skeletons.
Jaccard similarity was chosen as it is commonly used as an evaluation
metric in ACC, for example, to assess alignment between human- and
computer-identified sentence-type sets \cite{ref42}, inter-annotator
predicate overlap in logic clauses \cite{ref43}, and compare LLM-generated
Picat code against a human-verified gold standard \cite{ref14}. A signature
is a formal representation of a code component (i.e., a class or
function denoted as \(Sig( \bullet )\) in Equation~\ref{eq:structure-reward}) that also encodes
its nested position within the hierarchy.

The Jaccard similarity is the size of the intersection of the two
signature sets divided by the size of their union, yielding a score
between 0 and 1. For example, if the ground truth set of signatures is
\{A, B, C\} and the generated set is \{A, B, D\}, the intersection has
size 2 and the union size 4, resulting in a reward of \(+ 0.5\). This
evaluates the entire code skeleton, rewarding the model for identifying
both the right components and their correct placement. The accuracy
reward is defined as:

\begin{equation}
R_{\mathrm{structure}}(y,\widehat{y}) =
\begin{cases}
\dfrac{\bigl| \mathrm{Sig}(y) \cap \mathrm{Sig}(\widehat{y}) \bigr|}{\bigl| \mathrm{Sig}(y) \cup \mathrm{Sig}(\widehat{y}) \bigr|}, & \text{if } \mathrm{Sig}(y) \cup \mathrm{Sig}(\widehat{y}) \neq \varnothing \\[8pt]
0, & \text{otherwise}
\end{cases}
\label{eq:structure-reward}
\end{equation}

To ensure that the outputs are well-formed, two penalty-based
formatting rewards are included, as shown in Fig.~\ref{fig:3}a and Fig.~\ref{fig:3}b
respectively. The first is a syntactic reward \(R_{\text{syntax}}\),
which applies a penalty of \(- 1\) if \(\widehat{y}\) is not a
syntactically valid Python script according to the language
specification \(\mathcal{L}_{Python}\). This discourages the
model from producing malformed code that cannot be parsed into a
complete code skeleton. The syntactic reward function is defined as:

\begin{equation}
R_{\mathrm{syntax}}(\widehat{y}) =
\begin{cases}
-1, & \text{if } \widehat{y} \neq \varnothing \text{ and } \widehat{y} \notin \mathcal{L}_{\mathrm{Python}} \\[4pt]
0, & \text{otherwise}
\end{cases}
\label{eq:syntax-reward}
\end{equation}

The second is a length reward \(R_{\text{length}}\) designed to prevent
reward hacking. It applies a larger penalty of \(-3\) if \(\widehat{y}\)
is both shorter than a minimum length threshold of \(L_{\min}=100\)
tokens and deviates in length from \(y\) by more than a tolerance of
\(\tau=25\) tokens. These values are chosen heuristically to discourage
trivially short outputs while allowing short outputs close to the
reference length to remain unpenalized.
The magnitudes of the syntax and length rewards are
calibrated to the severity of the error. The \(- 1\) penalty for a
syntactic error is scaled to exactly negate the maximum possible score
of \(+ 1\), so an invalid script never yields a net positive reward,
while the larger \(- 3\) reflects that a trivially short output is a
more severe failure warranting a stronger deterrent. The length reward
function is defined as:

\begin{equation}
R_{\mathrm{length}}(y,\widehat{y}) =
\begin{cases}
-3, & \text{if } \lvert L(y) - L(\widehat{y}) \rvert > \tau \text{ and } L(\widehat{y}) < L_{\min} \\[4pt]
0, & \text{otherwise}
\end{cases}
\label{eq:length-reward}
\end{equation}

\section{Experimental setup}
\label{sec:experimental-setup}

This section describes the experimental setup used to evaluate P4IR.
The experiments used the dataset described in Section~\ref{sec:dataset},
comprising 664 training pairs and 68 test pairs. The following subsections
present the evaluation metrics, SFT and GRPO training settings, and
baseline comparison protocol.

\subsection{Evaluation metrics for code skeleton comparison}
\label{sec:evaluation-metrics}

This paper uses two metrics to evaluate the accuracy of the generated
code skeletons. Both serve as proxies for the remediation effort a rule
expert would need i.e., they quantify the editing operations needed to
align a generated skeleton with the ground truth. The first metric
captures differences in the skeleton's hierarchical structure, while
the second captures finer token-level differences such as naming.

\begin{figure}[ht]
\centering
\includegraphics[width=\linewidth]{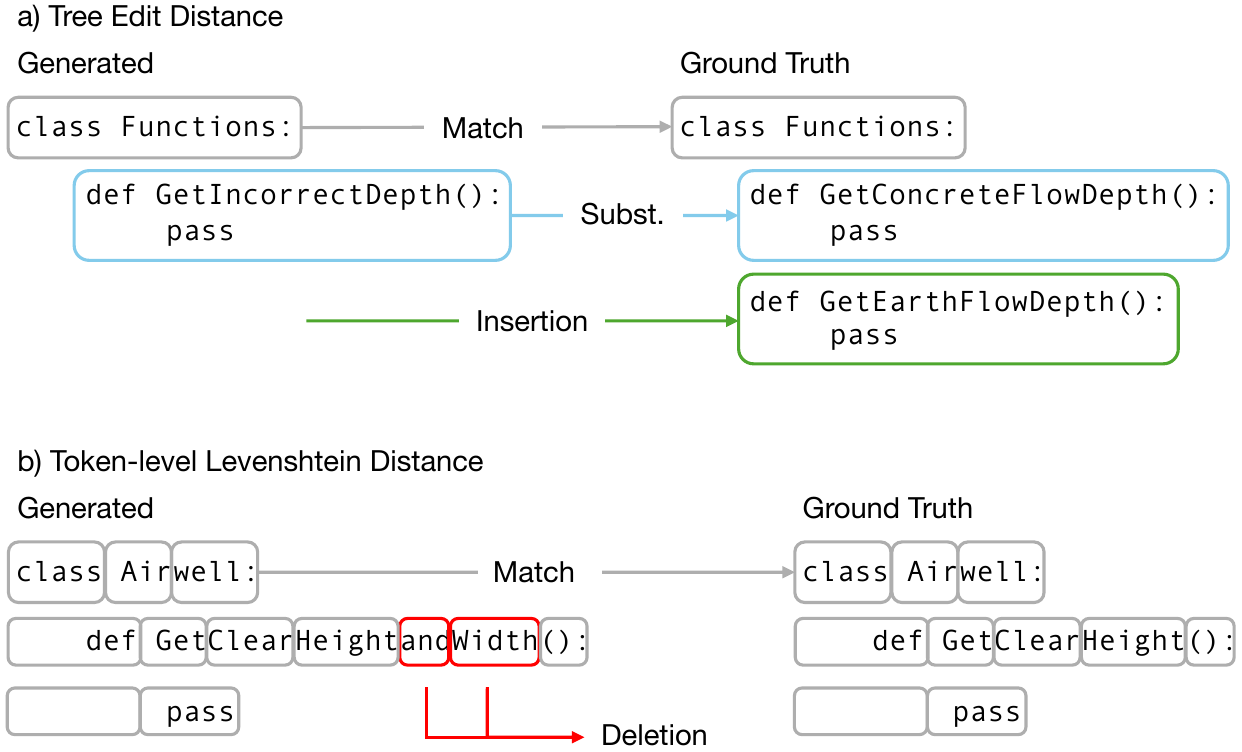}
\caption{Visualization of (a) tree edit distance and (b) token-level Levenshtein distance measures between generated and ground truth code skeletons.}
\label{fig:4}
\end{figure}

The first metric, as shown in Fig.~\ref{fig:4}a, quantifies the dissimilarity
between the structures of the generated and reference code skeletons,
such as incorrect nesting or missing class definitions. It is computed
in three steps. First, both code skeletons are parsed into simplified
Abstract Syntax Trees (ASTs) that retain the names and types of all class
and function definitions, together with their hierarchical relationships.
Next, each tree is represented as an ordered, labelled tree that preserves
parent-child relationships and sibling order. Finally, the Zhang--Shasha
algorithm \cite{zhang1989trees} is applied to compute the tree edit
distance between the generated and reference skeletons i.e., the minimum
number of node-level operations (insertions, deletions, and label
substitutions) required to transform one tree into the other, with each
operation assigned a unit cost.

To complement this, the token-level Levenshtein distance provides a more
granular comparison, as shown in Fig.~\ref{fig:4}b. While character-level
Levenshtein distance is common for natural language, a token-level
metric is more appropriate for source code, for three reasons. First,
the fundamental units of meaning in code are tokens (e.g., keywords,
identifiers, operators), not individual characters, so evaluating at the
token level better reflects the structure of the code. Second, it aligns
the evaluation with the model's generative process, since LLMs predict
sequences of tokens. Third, it is less sensitive to trivial variations
such as extra whitespace while still penalizing substantive errors.
To compute it,
the generated and reference skeletons are encoded into token sequences
using the model's tokenizer, and the standard Levenshtein distance is
applied to obtain the minimum number of single-token edits required to
transform one into the other. The two metrics evaluate the model's ability to generate accurate code
skeletons at both the structural and token levels.

\subsection{Experimental details for implementation and training}
\label{sec:training-settings}

Mistral 7B Instruct v0.3 was fine-tuned using LoRA for 14 epochs in
total, with checkpoints from epochs 4, 6, 8, 10, 12, and 14 selected as
initial policies for the subsequent GRPO stage. This range was chosen
as policies trained for fewer than 4 epochs were undertrained (i.e., a
further 1--2 epochs still produced large gains), while the performance
gains from SFT began to plateau after epoch 8, providing a suitable range
to study the interaction between SFT and GRPO.

All GRPO experiments were conducted on a single NVIDIA A100 40GB GPU.
Following Shao et al. \cite{ref12},
the learning rate was set to \(1e - 6\) and optimized using AdamW with
$\beta_1 = 0.9$ and $\beta_2 = 0.95$. A cosine learning rate scheduler
with a warmup ratio of 0.1 was employed. Training was performed with an
effective batch size of 4, achieved through 4 steps of gradient
accumulation.

For each prompt in a batch, the policy generated \(G = 16\) responses
with a high sampling temperature of 1.0 and a minimum probability
sampling of 0.1 \cite{ref44}, chosen to encourage diverse responses for
each prompt. The main experiments were run for a total of 2000 steps,
with model checkpoints saved every 200 steps. GRPO training was performed
with three random seeds for each initial SFT checkpoint. The KL divergence
coefficient \(\beta_{KL}\) was set to the default value of 0.04
\cite{ref12}. No hyperparameter tuning was performed i.e., standard values
were used to keep the focus on the method, leaving an exhaustive search
out of scope.

\subsection{Experimental details for comparison with baseline models}
\label{sec:baseline-settings}

To evaluate the efficacy of the framework against the SOTA LLMs, a
comparative analysis was conducted against five LLMs through their
respective APIs: Claude Opus 4.5, Claude Sonnet 4.5, GPT-5.2,
Qwen-3-Max, and GLM-4.7. These baseline models were selected to
represent the recent frontier of general-purpose LLMs.

\begin{figure}[ht]
\centering
\includegraphics[width=\linewidth]{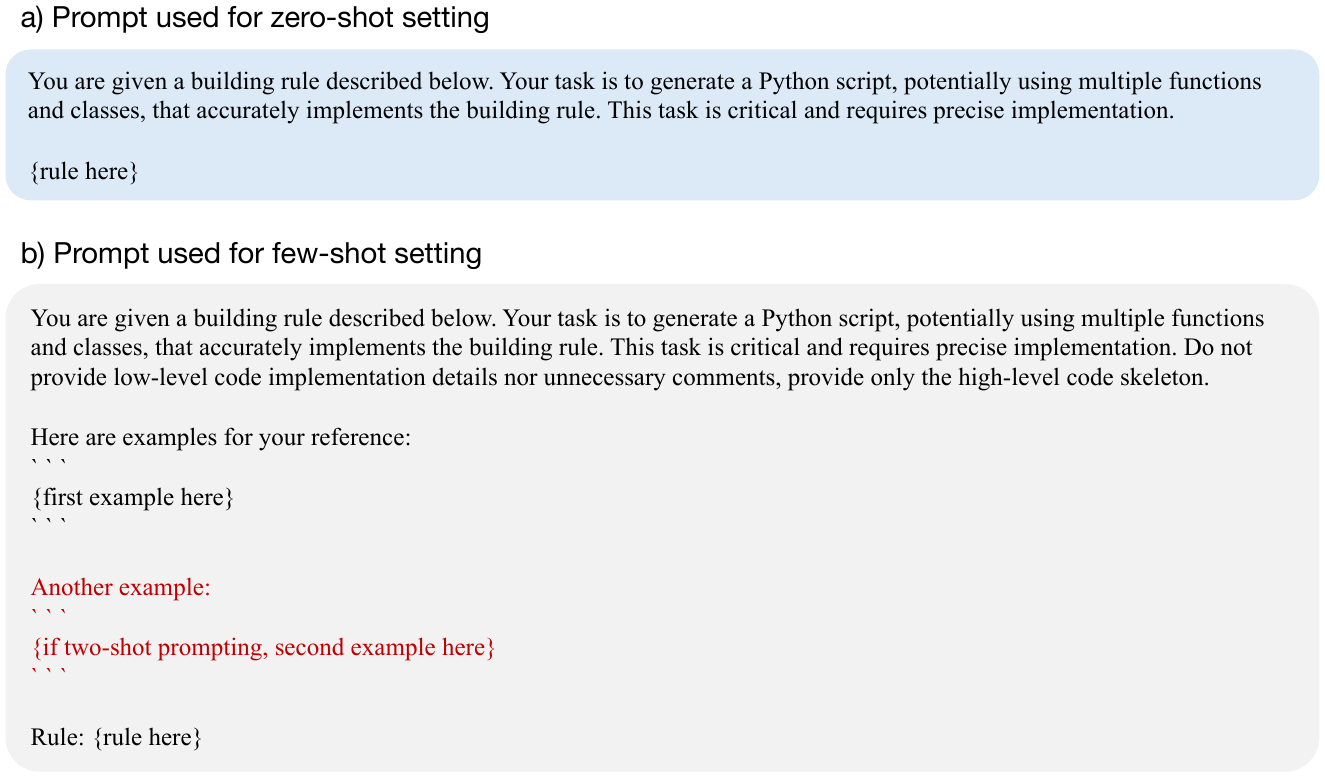}
\caption{Prompts relating to (a) zero-shot setting in the proposed framework and (b) few-shot setting in SOTA LLMs.}
\label{fig:5}
\end{figure}

Unlike the framework, which is evaluated in a zero-shot setting (i.e.,
without any examples during inference time, as shown in Fig.~\ref{fig:5}a), the
baseline models were evaluated using few-shot prompting. This simulates
a typical workflow where a general-purpose model is provided with
context to guide its generation. Specifically, for each test query, a
Retrieval-Augmented Generation (RAG) approach was used to retrieve
top-\emph{k} most similar code examples for that rule from the training
set, ranked by cosine similarity of their vector embeddings. In this
paper, one-shot and two-shot prompting (i.e., \(k = 1\) and \(k = 2\))
were tested. These retrieved examples were then inserted into the
original prompt, as shown in Fig.~\ref{fig:5}b.

Similar to the proposed framework, all baseline models were queried with
a temperature of 0.0 to maximize determinism and reproducibility for ACC
tasks. Aside from the insertion of few-shot examples, the baseline
prompts, as shown in Fig.~\ref{fig:5}b, contain slightly more explicit
instructions than P4IR's, as the baselines lack the domain-specific SFT
stage. This RAG-based few-shot setup was chosen as it has been shown to
benefit LLMs in ACC domain by providing more contextually relevant
answers \cite{ref45}. For this comparison, four metrics were used i.e., tree
edit distance and token-level Levenshtein distance described in Section~\ref{sec:evaluation-metrics}, plus Jaccard distance \cite{ref14,ref43} and CodeBERTScore \cite{zhou2023codebertscore},
covering both the structure and the semantics of the generated code
skeletons.

\section{Results and discussion}

The results show that GRPO's effectiveness depends strongly on the initial
policy from SFT, revealing an SFT window in which GRPO is most effective.
This window is where the model has sufficient domain-specific knowledge
from SFT to generate plausible structures while retaining the flexibility
required for refinement through reinforcement learning. While GRPO
generally improves the accuracy of the generated code skeletons, its
benefits are realized only when starting from a suitable SFT checkpoint.

\subsection{Interplay between SFT maturity and GRPO effectiveness}

\begin{figure}[ht]
\centering
\includegraphics[width=\linewidth]{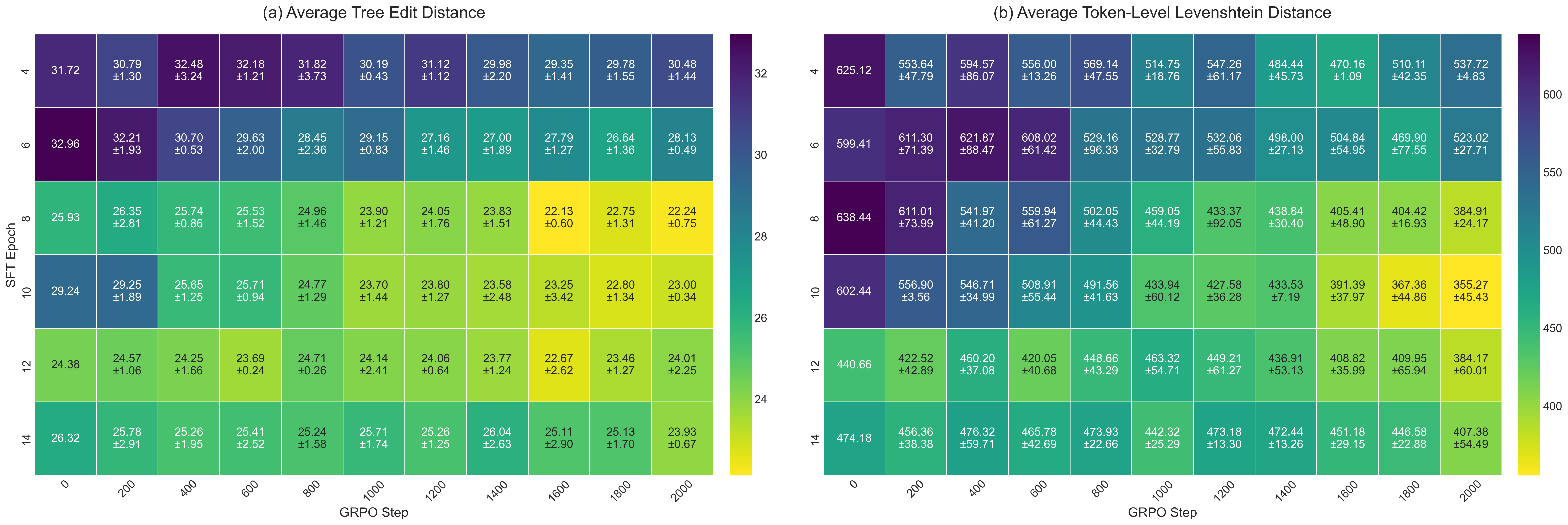}
\caption{(a) Average tree edit distance and (b) average token-level Levenshtein distance over 3 random seeds. Lower values indicate better performance.}
\label{fig:6}
\end{figure}

Fig.~\ref{fig:6} presents the primary evaluation metrics across GRPO optimization
steps on models from different SFT epochs. In both heatmaps, lower values
indicate a closer match to the ground truth (i.e., fewer edits to
transform the generated code skeleton into the reference). A clear
optimal performance window emerges for policies trained around 8 to 10
SFT epochs. For example, the epoch 10 policy, which begins with a tree
edit distance of 29.24, is progressively improved by GRPO to a minimum
of 23.00 after 2000 steps, a 21.3\% reduction in edit distance. A
similar trend is observed for the token-level Levenshtein distance,
which improves from 602.44 to a minimum of 355.27, a 41.0\% reduction.
In contrast, models at the extremes of SFT maturity show limited
benefit. Policies from earlier SFT epochs, such as epochs 4 and 6,
show weaker final performance. After 2000 GRPO steps, their tree edit
distances decreased from 31.72 to 30.48 and from 32.96 to 28.13
respectively. Despite these improvements, their final scores remained
higher than those of the epoch 8 and 10 policies, suggesting that less
mature SFT policies may constrain the accuracy attainable through
subsequent GRPO refinement. Conversely, policies from later checkpoints,
such as epochs 12 and 14, show limited benefit from GRPO as well.
While these models start with the strongest initial performance (i.e.,
a tree edit distance of 24.38 for epoch 12, which is the best score
achieved through SFT), they show negligible improvement from GRPO.
This indicates that the policy has become too deterministic, narrowing
its output distribution to the point where it lacks the exploratory
capacity for the GRPO reward signal to discover and reinforce better
code skeletons.

\subsubsection{Logit entropy across SFT epochs for model interpretability}

\begin{figure}[ht]
\centering
\includegraphics[width=\linewidth]{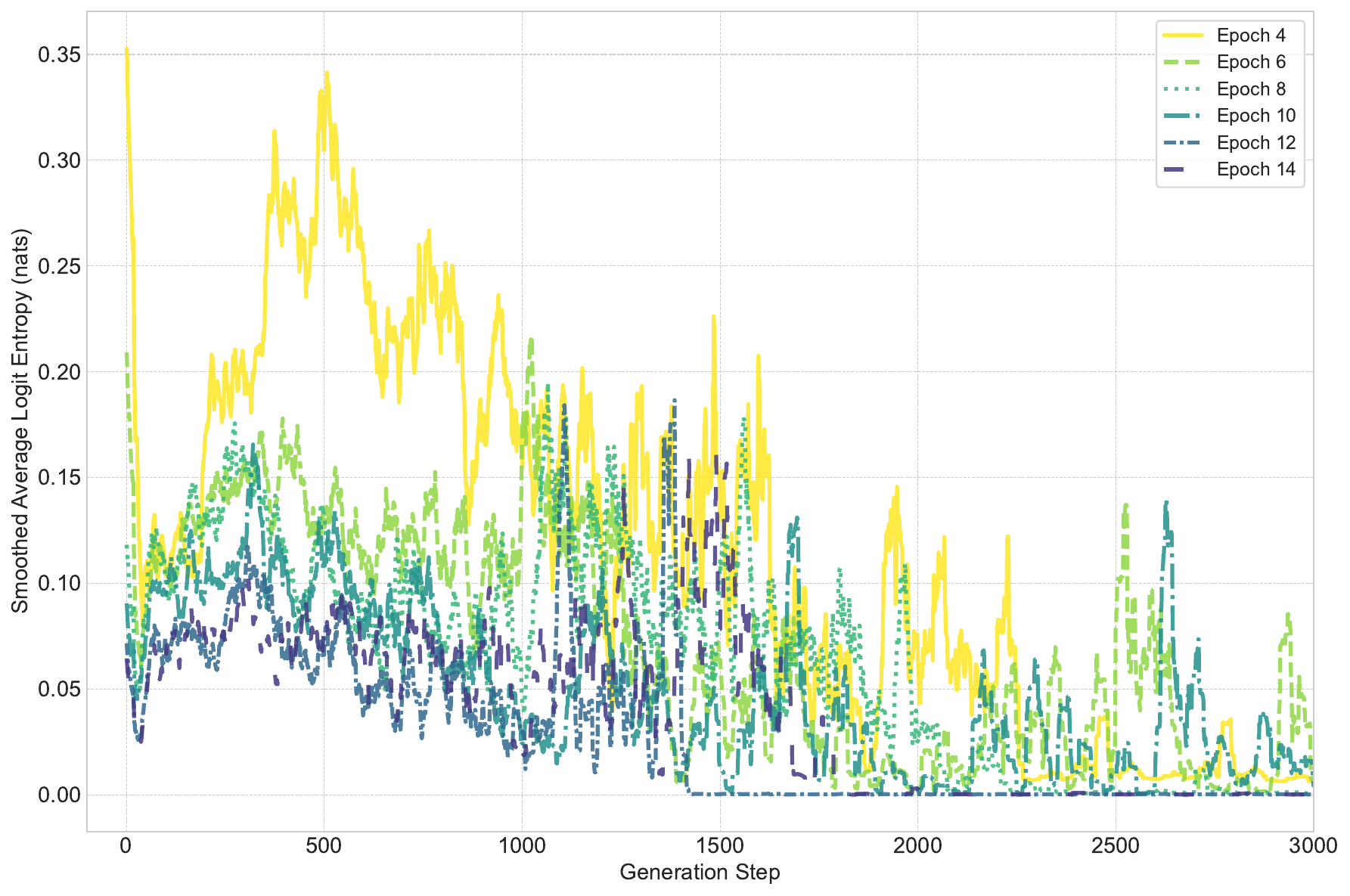}
\caption{Smoothed average logit entropy during generation.}
\label{fig:7}
\end{figure}

The underlying mechanism for this performance disparity can be
interpreted by analyzing the average logit entropy of the models during
generation, as shown in Fig.~\ref{fig:7}. First, the raw output logits of the model
are converted into a probability distribution \(P(x)\) over its entire
token vocabulary using the softmax function. The Shannon entropy \(H(X)\)
of this distribution is then computed using
\(H(X) = -\sum_{i} p(x_i)\log p(x_i)\).

In this context, logit entropy serves as a proxy for the predictive
uncertainty of the model, where high entropy indicates a more uniform
probability distribution over the next token, encouraging exploration,
while low entropy reflects high model confidence and leads to the
exploitation of patterns learned during SFT.

The analysis reveals three behavioral patterns that correspond to the
performance outcomes in Fig.~\ref{fig:6}. The epoch 4 policy exhibits consistently
high logit entropy, especially at the start, indicating that the model
is highly uncertain. This high degree of exploration leads to a wide
diversity of generated code skeletons, many of which may be incorrect
due to poor knowledge instillation during the SFT stage, making it
difficult for the reward signal to guide the policy toward a good
solution quickly enough. In contrast, the later-stage SFT policies from
epochs 12 and 14 show a rapid collapse in entropy, which approaches zero
after approximately 1400 generation steps. This policy collapse
indicates that the model has become overly deterministic, generating
nearly identical code skeletons. This lack of exploratory capacity
deprives the GRPO algorithm of diverse sampled outputs, preventing it
from discovering new outputs. The optimally performing policies from
epochs 8 and 10 maintain a moderate and stable level of entropy
throughout generation. This calibrated uncertainty may be well suited
to GRPO, balancing exploitation of known good structures against
exploration for better ones. This also suggests that monitoring logit
entropy may provide a useful interpretable heuristic \cite{ref46} for
identifying SFT checkpoints suitable for subsequent GRPO refinement.

\subsubsection{Effect of GRPO on semantics learned during SFT}

\begin{figure}[ht]
\centering
\includegraphics[width=\linewidth]{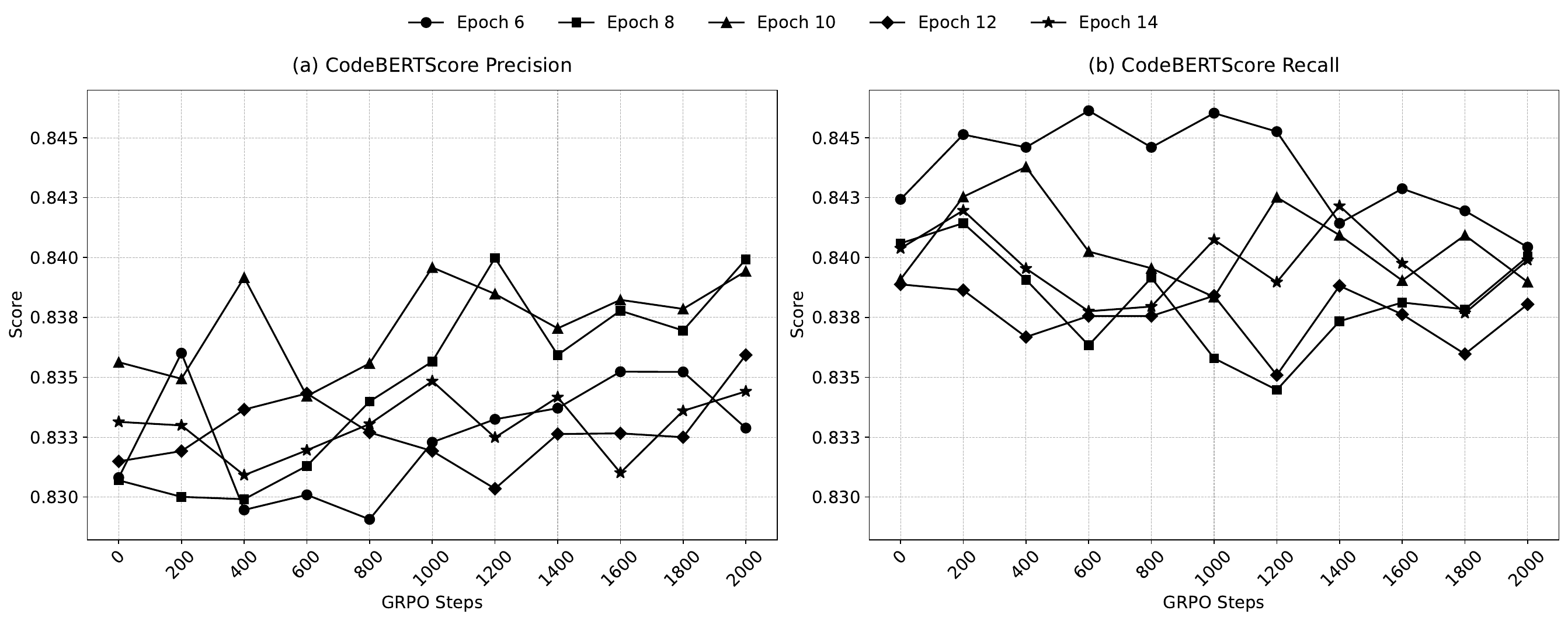}
\caption{(a) CodeBERTScore Precision and (b) CodeBERTScore Recall.}
\label{fig:8}
\end{figure}

While GRPO is effective at improving the accuracy of code skeletons, it
is essential to preserve the domain knowledge instilled during SFT. The
results indicate that it does not substantially alter the semantic
content learned during the SFT stage. Fig.~\ref{fig:8} shows the CodeBERTScore
precision and recall for different SFT epochs across the GRPO steps,
which measure the semantic similarity between the generated and
reference code skeletons.

Across all epochs, both CodeBERTScore precision and recall remain stable
throughout the GRPO process. For example, the precision for the epoch 10
model fluctuates within a narrow band from approximately 0.834 to
0.840, and its recall similarly stays between 0.838 and 0.843. Although
there is a general trend of slightly increasing precision and
decreasing recall as GRPO proceeds, the semantic understanding is
already established during SFT and is largely unchanged by GRPO. This
corroborates that SFT instills the domain-specific semantic knowledge
\cite{ref13}, whereas GRPO refines the structure of that knowledge to
improve accuracy without substantially changing its semantics.

\subsection{Mitigating model hallucinations}

The GRPO stage also reduces model hallucinations through the reward
functions. This analysis focuses on one measurable form of hallucination
i.e., the fabrication of extraneous classes or functions (false
positives, FP) not present in the ground truth. To quantify this, the
number of FP is measured in each generated output for a single training run.

An aggregated analysis across all epochs shows a reduction in these
hallucinations. The mean FP count decreased from 21.60 for SFT models to
18.60 after the GRPO stage, accompanied by a decrease in the standard
deviation from 23.87 to 18.00 and in the median FP count from 15.0 to
14.0, indicating that the GRPO models are more consistent as well as
more accurate on average. To verify the statistical significance of this
reduction, both non-parametric and parametric one-sided paired tests
were conducted. For these tests, FP counts were averaged across the six
checkpoints for each test regulation, separately for SFT and GRPO after
2,000 steps. This yielded 68 paired averages matched by regulation.
The Wilcoxon signed-rank test yielded \(p < 0.001\), corroborated by the
paired t-test with \(p = 0.011\), confirming a statistically significant
reduction in FP.

\subsection{Qualitative analysis of reference and generated code skeletons}

\begin{figure}[ht]
\centering
\includegraphics[width=\linewidth]{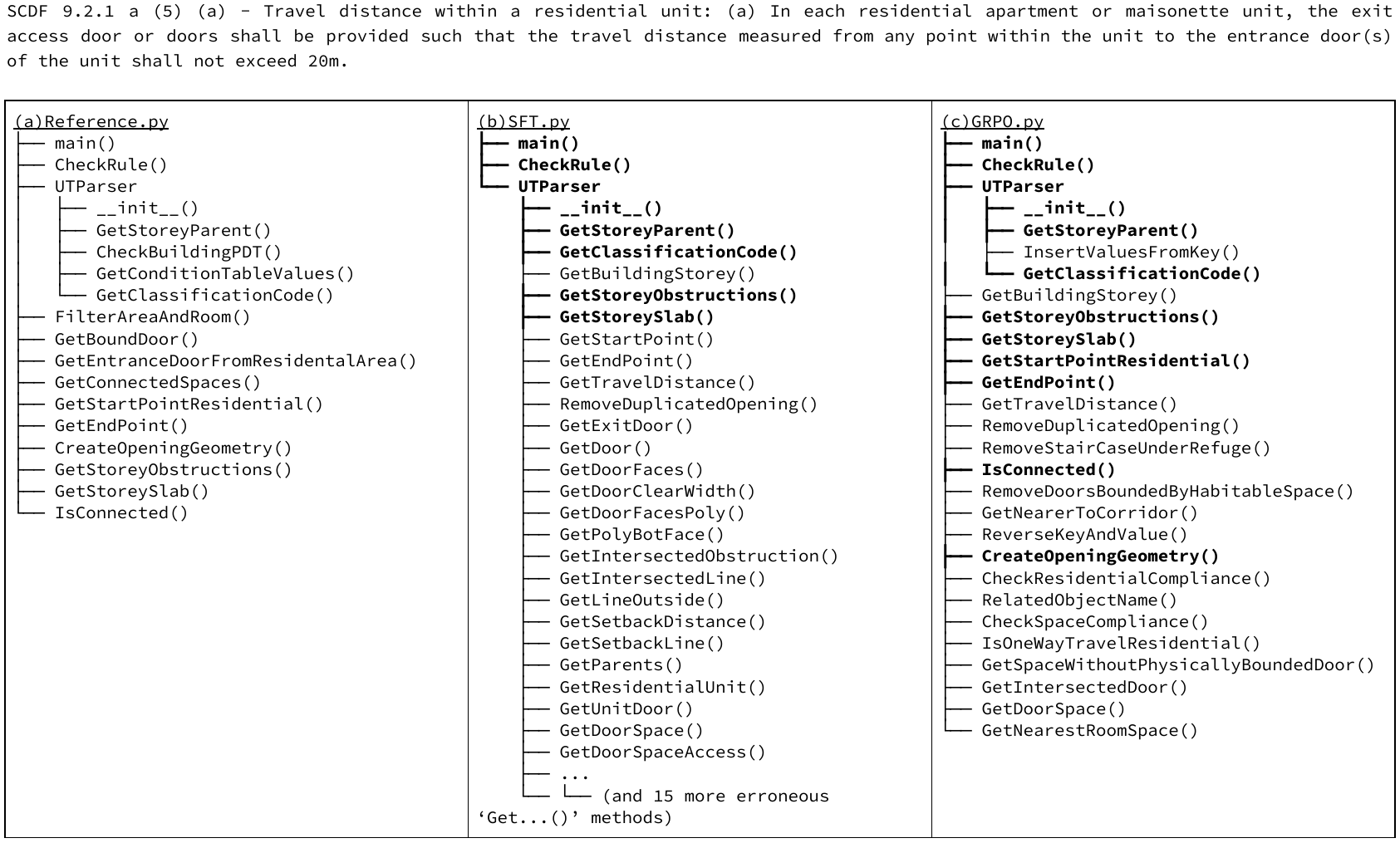}
\caption{Example pertaining to travel distance, showing (a) the reference, (b) the SFT-generated, and (c) the GRPO-refined code skeleton.}
\label{fig:9}
\end{figure}

\begin{figure}[ht]
\centering
\includegraphics[width=\linewidth]{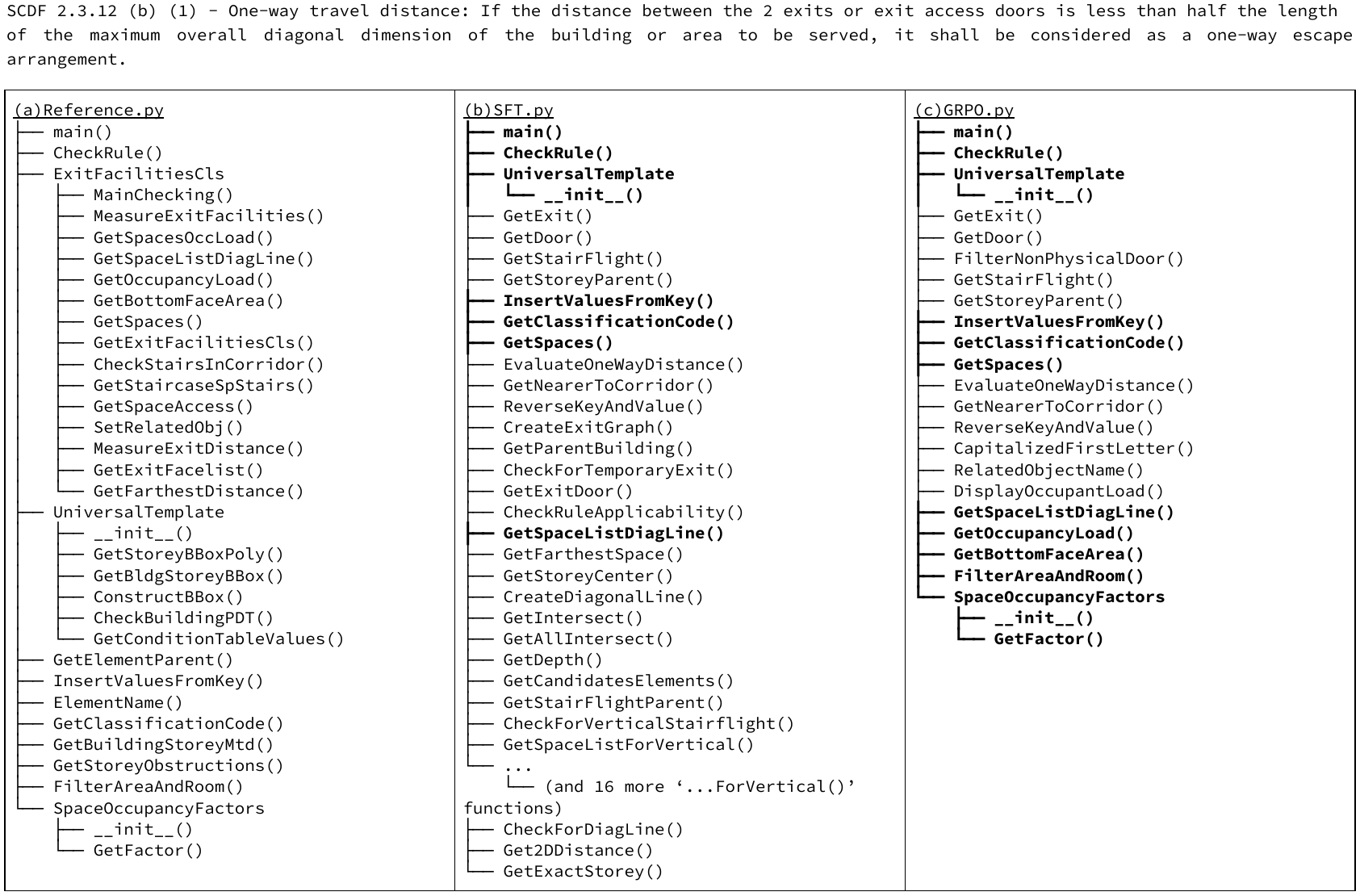}
\caption{Example pertaining to one-way escape arrangements, showing (a) the reference, (b) the SFT-generated, and (c) the GRPO-refined code skeleton.}
\label{fig:10}
\end{figure}

\begin{figure}[ht]
\centering
\includegraphics[width=\linewidth]{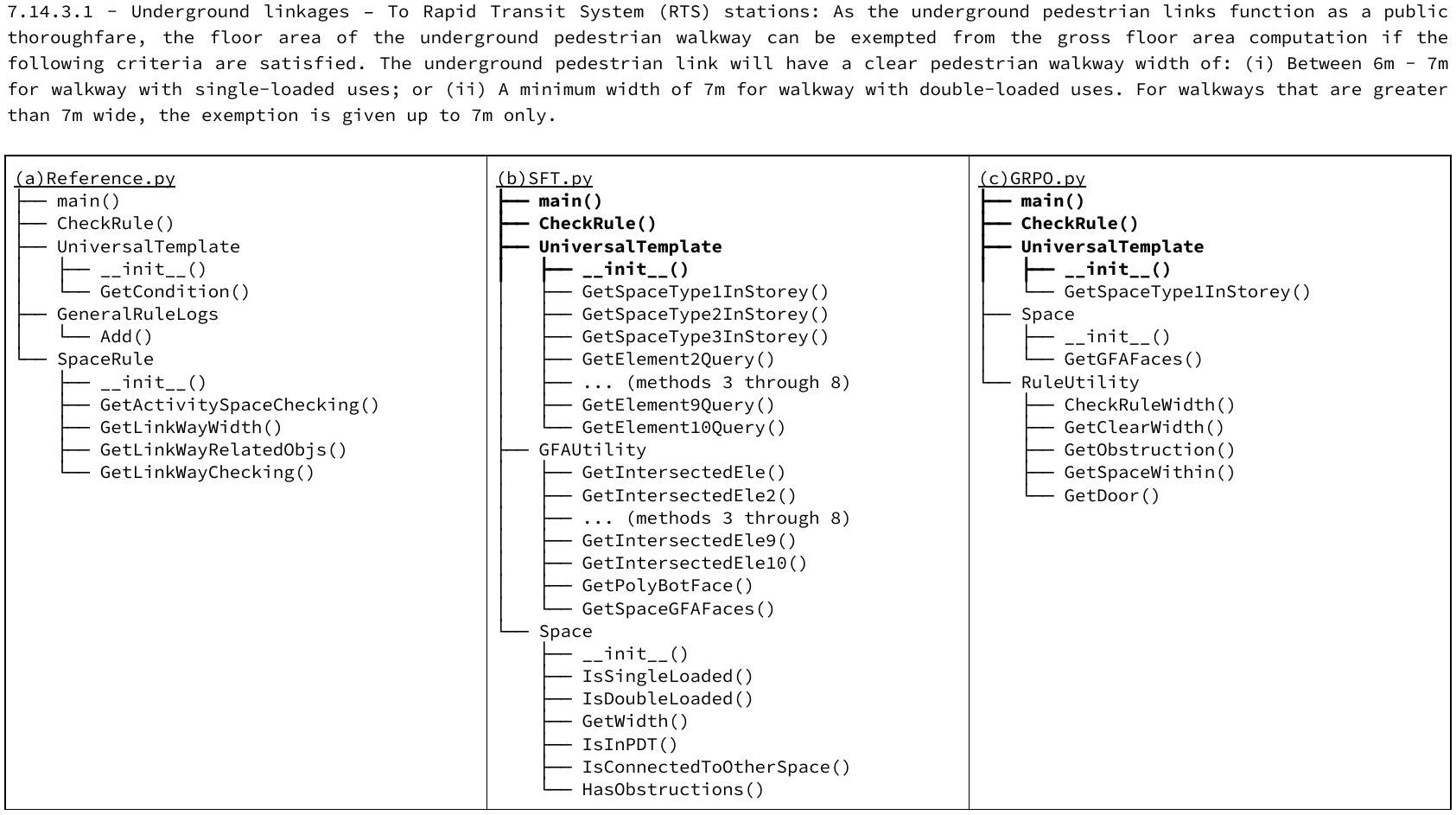}
\caption{Example pertaining to underground pedestrian linkages, showing (a) the reference, (b) the SFT-generated, and (c) the GRPO-refined code skeleton.}
\label{fig:11}
\end{figure}

In this section, three code skeletons are presented in Fig.~\ref{fig:9} to Fig.~\ref{fig:11}
as a qualitative comparison to further illustrate how GRPO improves
accuracy while reducing hallucinations. These examples compare the
initial SFT output with the refined code skeleton produced after the
GRPO stage. Definitions shown in bold correspond to those present in the
reference.

Fig.~\ref{fig:9} shows how GRPO reduces hallucinations and corrects the hierarchy.
The SFT model in Fig.~\ref{fig:9}b produces an inaccurate code skeleton. While the
SFT model identifies the \emph{`UTParser'} class, it proceeds to
incorrectly define all subsequent logic as methods under it, including
hallucinating over 15 erroneous \emph{`Get\ldots()'} methods. The model
after the GRPO stage corrects these issues. It removes the majority of
irrelevant hallucinated methods, then nests only the relevant methods
under \emph{`UTParser'} while terminating the list appropriately, so
that subsequent functions such as \emph{`GetStoreyObstructions()'} and
\emph{`GetStoreySlab()'} are correctly placed outside the
\emph{`UTParser'} class definition. This demonstrates an improved
understanding of object-oriented code structure.

Fig.~\ref{fig:10} further highlights the strength of the framework. Apart from
reducing hallucinations as in the previous example, this case shows GRPO
recovering entire class definitions and correct methods absent from the
SFT output. The SFT model shown in Fig.~\ref{fig:10}b produces a largely unusable
code skeleton, overwhelmed by a long list of hallucinated methods. In
contrast, the GRPO model shown in Fig.~\ref{fig:10}c reduces this noise and
improves completeness. It regenerated the complete
\emph{`SpaceOccupancyFactors'} class, including its
\emph{`\_\_init\_\_()'} and \emph{`GetFactor()'} methods, a key
abstraction for this regulation that the SFT model had failed to
produce, and recovers other methods like \emph{`GetOccupancyLoad()'} and
\emph{`GetBottomFaceArea()'}.

Fig.~\ref{fig:11} shows how GRPO improves coherence by correcting structural flaws
and removing certain hallucinations. The SFT model shown in Fig.~\ref{fig:11}b
produces an inefficient structure, creating a \emph{`GFAUtility'} class
populated with a long series of incrementally named and hallucinated
methods (e.g., \emph{`GetIntersectedEle2'} through
\emph{`GetIntersectedEle10'}). The GRPO stage removes these redundant
methods and reorganizes the logic into a more sensible
\emph{`RuleUtility'} class. This refinement also illustrates semantic
equivalence. While the GRPO-generated methods, such as
\emph{`CheckRuleWidth'}, are not syntactically identical to the
\emph{`GetLinkwayWidth'} method in the reference output, they capture
the same functional intent. This is a limitation for precise
replication, but also shows the model generating accurate and equivalent
abstractions.

The improvements illustrated in these examples were obtained using the
reward system described in Section~\ref{sec:rewards}. Removing the syntax and length
rewards led to a model collapse after approximately 1000 GRPO steps,
producing syntactically invalid definitions (e.g.,
\texttt{defcheck\_door\_width(width):} instead of
\texttt{def check\_door\_width(width):}) and excessively short outputs
consisting of a single function containing only a \texttt{pass}
statement. Without the Jaccard reward, performance initially remained
close to the starting SFT checkpoint, with no clear improvement, and
subsequently deteriorated gradually.

\subsection{Comparative results of P4IR against SOTA models}

To position the proposed framework within the landscape of current LLMs,
the optimal policy was evaluated against five SOTA models via few-shot
prompting: Claude Opus 4.5, Claude Sonnet 4.5, GPT-5.2, Qwen-3-Max, and
GLM-4.7, as described in Section~\ref{sec:baseline-settings}. As summarized in Table~\ref{tab:comparison}, P4IR
outperforms all five baselines on every metric.

\begin{table}[t]
\caption{Comparative performance of P4IR (zero-shot on epoch 8 model) against retrieval-augmented SOTA models via few-shot prompting. Downward arrow indicates lower is better, and vice versa. The FT + GRPO is computed across 3 random seeds.}
\label{tab:comparison}
\begin{center}
\small
\setlength{\tabcolsep}{4pt}
\resizebox{\linewidth}{!}{%
\begin{tabular}{llccccc}
\toprule
\textbf{Model} & \textbf{Setting} & \shortstack{\textbf{Tree Edit}\\\textbf{Distance} ($\downarrow$)} & \shortstack{\textbf{Token-level}\\\textbf{Levenshtein}\\\textbf{Distance} ($\downarrow$)} & \shortstack{\textbf{Jaccard}\\\textbf{Distance} ($\downarrow$)} & \shortstack{\textbf{CodeBERT-}\\\textbf{Precision} ($\uparrow$)} & \shortstack{\textbf{CodeBERT-}\\\textbf{Recall} ($\uparrow$)} \\
\midrule
\multirow{2}{*}{Claude Opus 4.5} & 1-shot & 34.46 & 808.94 & 0.9242 & 0.7579 & 0.8004 \\
                                 & 2-shot & 37.97 & 802.62 & 0.9178 & 0.7689 & 0.8100 \\
\addlinespace
\multirow{2}{*}{Claude Sonnet 4.5} & 1-shot & 29.40 & 607.75 & 0.9087 & 0.7666 & 0.7962 \\
                                   & 2-shot & 31.54 & 594.34 & 0.8974 & 0.7805 & 0.8045 \\
\addlinespace
\multirow{2}{*}{GPT-5.2} & 1-shot & 30.00 & 1234.22 & 0.9556 & 0.7054 & 0.7800 \\
                         & 2-shot & 31.37 & 1283.87 & 0.9515 & 0.7142 & 0.7929 \\
\addlinespace
\multirow{2}{*}{Qwen-3-Max} & 1-shot & 24.15 & 411.21 & 0.8808 & 0.7784 & 0.7922 \\
                            & 2-shot & 24.21 & 395.13 & 0.8868 & 0.7833 & 0.7878 \\
\addlinespace
\multirow{2}{*}{GLM-4.7} & 1-shot & 25.99 & 546.53 & 0.8806 & 0.7815 & 0.7971 \\
                         & 2-shot & 23.68 & 475.78 & 0.8721 & 0.7884 & 0.7915 \\
\midrule
\multirow{3}{*}{P4IR} & FT only & 25.93 & 638.44 & 0.7530 & 0.8307 & \textbf{0.8406} \\
                     & FT + GRPO & \textbf{22.24} & \textbf{384.91} & \textbf{0.7081} & \textbf{0.8395} & 
                     0.8396 \\
                     & & $\pm0.75$ & $\pm24.17$ & $\pm0.0047$ & $\pm0.0029$ & $\pm0.0021$ \\
\bottomrule
\end{tabular}}
\end{center}
\end{table}

In general, Claude Opus 4.5, Claude Sonnet 4.5, and GPT-5.2 perform
poorly across all metrics. The 1-shot versus 2-shot comparison also
reveals a counter-intuitive trend for these three models. In fact,
increasing the number of exemplars from one to two degraded the tree
edit distance (e.g., Claude Opus worsened from 34.46 to 37.97). This
context pollution suggests that for highly specialized tasks like ACC,
simply injecting more examples into the context window of a
general-purpose model may be insufficient and potentially detrimental.
The additional context may introduce conflicting patterns that confuse
the model. In contrast, P4IR internalizes domain knowledge through SFT
and refines it via GRPO, yielding a more stable policy that does not
rely on inference-time retrieval or external vector databases.

Qwen-3-Max and GLM-4.7 exhibit a distinct performance pattern compared
to the weaker baselines. They make better use of few-shot prompting,
achieving competitive tree edit and token-level Levenshtein distance.
For example, with two-shot prompting, Qwen-3-Max and GLM-4.7 achieved
tree edit distances of 24.21 and 23.68, and token-level Levenshtein
distances of 395.13 and 475.78, respectively. However, their agreement
with the reference class and function identifiers remains weaker than
that of P4IR. Despite lower edit distances, these two models, like all
other baselines, recorded high Jaccard distances (i.e., \(>0.87\)), which
is worse than that of P4IR. These scores should be read in light of how
strict the Jaccard metric is for ACC, since it penalizes even minor
lexical variation in predicate naming. As a reference point, Xue and
Zhang \cite{ref43} reported that even between human domain experts, the
inter-annotator Jaccard distance is approximately 0.63, indicating that
the reference is only one of several valid structures, and that a
distance near zero is neither achievable nor expected. While this
figure measures a related rather than identical quantity, it provides
a useful floor. P4IR's 0.7081 sits close to this expert-disagreement
level, whereas the SOTA models (i.e., around 0.90) remain well above it,
indicating a more fundamental failure to recover the correct
domain-specific class and function identifiers.

The CodeBERTScores show a similar pattern, where the baseline scores
lower on semantic similarity, indicating that even Qwen-3-Max and
GLM-4.7, which mimic the code skeleton better than other baselines, fail
to generate the correct domain-specific identifiers. P4IR achieves the
highest CodeBERTScore precision and recall of 0.84. While the baselines
achieved reasonable recall (i.e., peaking at 0.81 for Claude Opus 4.5),
their precision scores remained lower than those of P4IR. In ACC
systems, this lower precision may increase the need for expert review
of the generated code skeletons. These results indicate that P4IR
produces code skeletons with greater structural and semantic similarity
to the references under the evaluated metrics.

\section{Limitations and future work}

While P4IR improves the accuracy of code skeletons and reduces
hallucinations, several limitations qualify these findings and highlight
directions for future work. First, P4IR rewards structural accuracy
rather than functional correctness. The evaluation also relies on
reference skeletons and does not independently establish whether the
generated components correctly and completely represent the regulatory
requirements. Future work could develop an expert assessment procedure
to evaluate these components directly against the source regulations.
Although the accuracy of the IRs bridging building regulations and the
computer-processable rules is improved, whether the model can be
reinforced directly for functional correctness remains an open question.
Future work could explore rewarding the model based on the outcomes of
BIM unit tests (i.e., pass/fail results from a BIM sandbox environment)
corresponding to the generated computer-processable rules. While this
would more directly optimize for executable accuracy, it presents
challenges, including the need for extensive test-case datasets and the
added complexity of creating rule interpretation datasets.

Second, this paper focuses on a single form of IR (i.e., code skeleton),
leaving its generalizability to other structured formats unexplored.
Building regulations also differ widely across countries in structure
and terminology, leading to semantic inconsistencies in vocabulary. P4IR
has not been tested on this variation due to the scarcity of labeled
rule interpretation datasets, so its ability to generalize without
country-specific training remains unknown. Future work could investigate
additional specialized reward functions and this GRPO-based approach on
other IRs, such as atomic functions or ontologies, to assess its broader
utility in ACC. P4IR could also be evaluated on schema-constrained IRs
(e.g., LegalRuleML), where fixed syntax reduces stylistic variation and
may yield a more stable target for reward design and evaluation.

\section{Conclusion}

This paper introduced the P4IR framework, an approach to improve the
accuracy of code skeletons while reducing hallucinations. The framework
first used SFT to instill foundational domain knowledge before applying
GRPO to refine the generated code skeletons. The results showed that
P4IR reduced the distance between the generated and ground-truth code
skeletons. Hallucinations in the form of redundant or fabricated classes
and functions were also reduced, contributing to more consistent outputs.

The comparison with frontier LLMs showed that, despite their recent
advances, these models lacked the capabilities for off-the-shelf
application to specialized, real-world ACC tasks. In contrast, P4IR
produced a specialized model that outperformed far larger baselines
while offering a privacy-preserving workflow, processing sensitive
regulatory and design data locally without compromising performance.
The findings demonstrate that applying GRPO to the initial SFT models
improves the accuracy of generated IRs while reducing hallucinated
classes and functions. They also show that additional SFT does not
necessarily produce a better starting policy for GRPO, providing
empirical guidance on combining the two training stages.

In practice, more accurate IRs with fewer hallucinations may reduce the
effort required to develop complete compliance-checking scripts. Future
work should assess this potential benefit through expert editing
studies and evaluate the functional correctness and reliability of the
completed scripts through execution-based testing. As the AEC industry
adopts automated workflows, P4IR offers a promising direction for
translating building regulations into computer-processable rules.

\end{document}